\documentclass[11pt,oneside,letterpaper]{article}
\pdfoutput=1
\usepackage[utf8]{inputenc}
\usepackage{amsmath,amssymb,latexsym}
\usepackage{slashed}
\usepackage{epstopdf}
\usepackage[dvips]{graphicx}
\usepackage{setspace}
\usepackage{mathtools}
\usepackage{amsfonts}
\usepackage{fancyhdr}
\usepackage{xcolor}
\usepackage{graphicx}
\usepackage{rotating}
\usepackage{comment}
\usepackage{color}
\usepackage{subcaption}
\usepackage[percent]{overpic}
\usepackage{cite}
\usepackage{braket}
\definecolor{darkgreen}{rgb}{0,0.5,0}
\definecolor{darkblue}{rgb}{0,0,0.6}
\definecolor{purple}{rgb}{0.4,.2,0.7}

\newcommand{\f}{\frac}

\newcommand{\be}{\begin{equation}}
\newcommand{\ee}{\end{equation}}

\usepackage[colorlinks=true,citecolor=darkgreen,linkcolor=black,urlcolor=purple]{hyperref}
\usepackage[normalem]{ulem}
\usepackage{pdfsync}
\usepackage{xcolor}
\def \sec{\begin{section}}
\def \esec{\end{section}}

\usepackage[margin=0.8in]{geometry}

\renewcommand{\tilde}{\widetilde}

\def \la {\lambda}

\def \ep {\epsilon}
\def \th {\theta}

\def \si {\sigma}

\def \Qtt {\tilde{Q}}

\def \pr {\partial}
\def \ra {\rightarrow}

\def \beq { \begin{equation}}
\def \eeq {\end{equation}}

\def \at {\biggl{\vert}}

\DeclareMathOperator*{\Tr}{Tr}

\def \l {\left(}
\def \r {\right)}

\def \bra {\langle}
\def \ket {\rangle}

\begin{document}

\onehalfspacing

\begin{center}

~
\vskip5mm

{\LARGE  {
Bra-ket wormholes and Casimir entropy  \\
}}

\vskip10mm

Alexey Milekhin and Amirhossein Tajdini

\vskip5mm

{\it Department of Physics, University of California, Santa Barbara, CA 93106, USA
} 

\vskip5mm
{\tt  milekhin@ucsb.edu, ahtajdini@ucsb.edu}

\end{center}

\vspace{4mm}

\begin{abstract}
\noindent

Bra-ket wormholes are non-trivial saddles in Euclidean gravity. They have to be sustained by negative Casimir energy of matter fields inside the throat. However, Casimir energy is very sensitive to boundary conditions and in presence of gauge symmetries one has to integrate over all possible boundary conditions for the matter fields, as they are a part of bra-ket wormhole moduli. For non-Abelian gauge groups the corresponding measure for the boundary conditions, which we call Casimir entropy, is non-trivial and it competes with the Casimir energy.
We find that for large gauge groups this significantly affects the bra-ket wormhole action and modifies the phase diagram. Despite that, we do not find any violations of strong subadditivity in the setup proposed by Chen, Gorbenko and Maldacena.

 \end{abstract}
%\vspace{.2in}
%\vspace{.3in}

\pagebreak
\pagestyle{plain}

\setcounter{tocdepth}{2}
{}
\vfill
\tableofcontents

\newpage

\section{Introduction}

\begin{figure}

    \centering
    \includegraphics[scale=0.8]{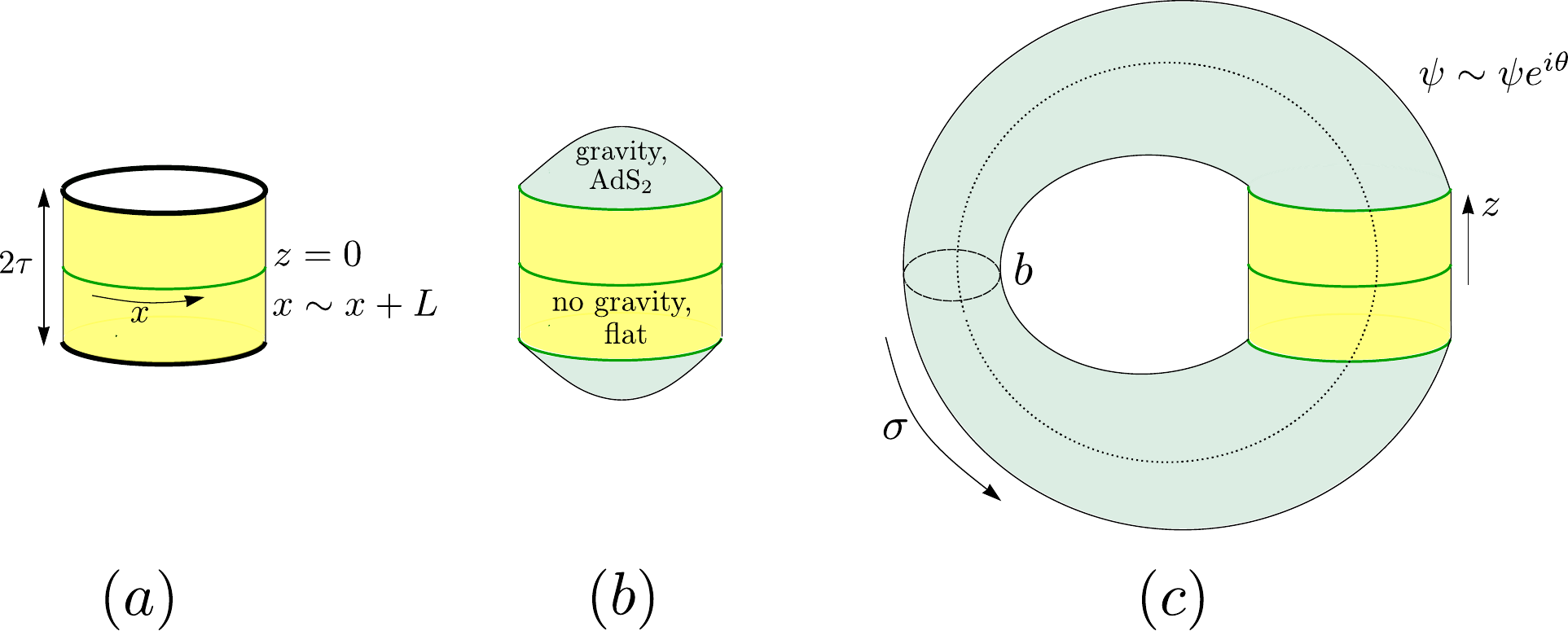}
    \caption{(a) \label{fig1} We use Euclidean path integral to prepare a state of a non-gravitational system at $z=0$ and then compute the overlap with itself. We use a holographic system (black circle) as a boundary condition. In the dual gravity picture, we have to sum over all geometries (green) ending on the red circle. (b) One possible saddle: two disconnected discs. (c) Bra-ket wormhole saddle. Notice that it contains an extra non-contractable cycle along $z$ direction. Matter fields (denoted $\psi$) charged under a gauge symmetry can undergo a non-trivial transformation when they travel along this closed cycle. Parameter $\theta$ can be transformed into a flat gauge connection. We have to integrate over all possible $\theta$. }
    \label{fig:drawing}
\end{figure}
Gravitational wormholes are fascinating objects as they represent space-times with non-trivial topologies. Recently there was a lot of progress in demonstrating the unitarity of black hole evaporation by including the wormhole contributions \cite{Penington, Almheiri:2019psf, east_coast, west_coast}. However, they still pose a lot of puzzles, as they potentially clash with factorization and often are not true saddles of the action \cite{saad2018semiclassical,Cotler:2020lxj,Cotler:2020ugk,Cotler:2021cqa}. 

Sometime ago Chen, Gorbenko and Maldacena discovered a new class of wormholes by studying the following setup \cite{Chen:2020tes}. We prepare a state of a non-gravitational theory (say 1+1 CFT) by using a holographic theory (say 1+0 Sachdev--Ye--Kitaev model \cite{SachdevYe,KitaevTalks,ms,Polchinski:2016xgd}) as a boundary condition - Figure \ref{fig:drawing}. We work in the Euclidean and place Sachdev--Ye--Kitaev (SYK) dot at $z=-\tau$ and consider its Euclidean evolution along the $x$ direction. 
In the dual gravity picture \cite{cft_breaking,Yang:2018gdb,Saad:2019lba} we will have to integrate over all possible Euclidean geometries in Jackiw--Teitelboim (JT) gravity (shown in green). We will assume that the matter CFT has transparent boundary conditions at $z=\pm \tau$ so that it can propagate into the gravity region. Two such saddles are shown in Figure \ref{fig:drawing} (b) and (c). 
Saddle $(b)$ is simple: it is just two Poincar\'e discs. Saddle $(c)$ is what is called bra-ket wormhole \cite{Chen:2020tes}.
It has to be sustained by negative Casimir energy coming from CFT matter. Direction $z$ is closed, so if we treat it as a spacial slice, a generic QFT will have a non-zero energy on that slice. Strictly speaking, it becomes Casimir energy only if we analytically continue (the perpendicular) angular $x$ direction to Lorentzian.

Two-disk saddle has no moduli (internal parameters).
Bra-ket wormhole has two non-trivial geometric moduli: throat size $b$ and a relative twist $\chi$ along the $x$ direction (not shown). One has to integrate over them. Integration over $\chi$ does not lead to any significant quantitative effects, it gives a subleading contribution to the action \cite{Chen:2020tes}. Integration over $b$ is dominated by a saddle point: gravity causes $b$ to shrink, but negative energy counteracts that. Depending on $L$ (circumference of the cylinder), $\tau$ and JT gravity parameters, either it is two disks which dominate the path integral or it is bra-ket wormhole.

Path integral instructs us to integrate over all possible moduli. Global symmetries do not imply any additional moduli.
In this paper we consider the case when we have gauge symmetries.   Notice that bra-ket wormhole contains an extra emergent closed cycle along $z$ direction - Figure \ref{fig:drawing} (c), dotted curve. If matter CFT is charged under a gauge symmetry it means that it lives in a certain fiber bundle over the 2d geometry. This fiber bundle might have a non-trivial connection which should be treated as moduli. In simple terms, CFT matter can undergo a non-trivial transformation which we schematically denote as $\psi \ra \psi e^{i\theta}$ when it travels along bra-ket wormhole cycle (Figure \ref{fig:drawing} (c), dotted line). Since $\theta$ is part of the moduli, we have to integrate over it. However, $\theta$ affects Casimir energy. For example, for free fermions the Casimir energy is minimal (and negative) when all fermions have antiperiodic boundary conditions. But this is just a single point in the moduli space. Moreover, for non-Abelian groups $\theta$ has non-trivial measure (e.g. Haar measure for $U(N)$).  We find that the measure for $\theta$, which we naturally call Casimir entropy, competes with the Casimir energy.
We call it entropy because, like entropy, it measures the number of states.
Interestingly, the most "entropic" configuration has zero Casimir energy:
\beq
\text{max measure}(\theta) \ra E_{\rm Casimir}=0.
\eeq
so it is "harder" for the bra-ket wormhole to exist when gauge symmetries are present.
It turns out this effect significantly modifies the phase diagram "two disks vs bra-ket wormhole" for large non-Abelian groups. 
Specifically we will work with $N$ free complex Dirac fermions in the fundamental representation of $U(N)$ gauge group. Casimir energy is proportional to $N$, whereas Casimir entropy (Haar measure for $U(N)$) is proportional to $N^2$, this is why it is an important effect for large $N$.
For generic non-zero gauge coupling this is a complicated computation so we will study the case of zero coupling. It means that we will care only about flat connections. They are equivalent to a twist-$\theta$ boundary condition. 
As a side note, it is much easier to consider Abelian gauge groups, such as $U(1)^N$ and $(\mathbb{Z}_2)^N$, as the corresponding measure for $\theta$ is flat (no Casimir entropy). We did not find any interesting quantitative effects in this case, so we would not present these computations here.

Bra-ket wormholes are important because they 
resolve a strong subadditivity (SSA) paradox \cite{Chen:2020tes}.
One can study matter entanglement entropy for
the two-disk saddle using island prescription \cite{Penington, Almheiri:2019psf, east_coast, west_coast}.
It turns out that the answers are not compatible with SSA of entanglement in a certain regime of parameters. The resolution is that in this regime it is bra-ket wormhole which dominates. Given that in our setup there is a big correction from Casimir entropy it is natural to ask if we still avoid SSA or not. The answer turns out to be positive, as long as JT entropy $S_0$ satisfies the inequality
\beq
S_0 \gtrsim \frac{N}{6} \log N.
\eeq
This inequality is pretty natural because we couple $N$ fermions to SYK dot with transparent boundary conditions so we should expect that we cannot "stuff" inside more than $S_0$ fermionic species. Our direct calculation shows that there is a logarithmic correction to this logic, the $\log N$ factor above. This can be viewed as 2d analog of the "species bound" of higher-dimensional quantum gravity \cite{universal,Casini:2008cr,Susskind:1994sm,Veneziano:2001ah,Dvali:2007hz,Dvali:2007wp}. 

Unfortunately, one "unrealistic" aspect of our setup is that 
we generally do not expect large gauge groups in the bulk. However, the notion of Casimir entropy we introduce here is pretty general for gravity computations with bra-ket wormholes. There might be other setups where Casimir entropy matters even for smaller gauge groups.

Another well--known aspect of quantum gravity is that it is incompatible with global symmetries \cite{abbott1989wormholes,gilbert1989wormhole,Kallosh:1995hi,Maldacena_2004,Milekhin:2021lmq,Chen:2020ojn,Hsin:2020mfa}. 
We will see that bra-ket wormholes also show inconsistency if we have a global symmetry instead of a gauge symmetry. In Section \ref{sec:paradox} we will demonstrate that it clashes with the purity of the prepared state. If a symmetry is gauged, integration over $\theta$ prohibits charged particles from going through the wormhole and the paradox is avoided. This is obvious if the symmetry is gauged everywhere. However, we should expect that if it is gauged in the gravity region only (in the non-gravitational region it is a global symmetry), then the conclusion must be the same. It is indeed the case, but this setup will require a careful analysis of the boundary conditions for the gauge connection. We will do it in Section \ref{sec:only} and show that one has to integrate over $\theta$ even if the symmetry is gauged only in the gravity region.

The rest of the paper is organized as follows. In Section \ref{sec:only} we analyze the case when non-gravitational region has a global symmetry which is gauged in the gravitational region. In Section \ref{sec:paradox} we discuss a purity paradox which arises when there is
a global symmetry everywhere. Section \ref{sec:casimir} discusses the Casimir energy and the origin of Casimir entropy. Section \ref{sec:braket} is the main section of this paper where we discuss the competition between Casimir energy and Casimir entropy and how it affects the
existence of bra-ket wormholes. The computation will reduce to a certain matrix model integral which we analyze in detail in the Appendix \ref{app:int}. We will conclude with analyzing SSA paradox in Section \ref{sec:ssa}.

\section{Gauge symmetry in the gravity region only}
\label{sec:only}
Consider a bra-ket wormhole saddle.
If a symmetry is gauged everywhere, both in the gravitational and non-gravitational regions, it is obvious that the connected saddle has an extra closed cycle and the corresponding gauge connection $A_\mu$ has a gauge-invariant holonomy:
\beq
\oint A_\mu dx^\mu,
\eeq
But what happens if the symmetry is gauged in the gravity region only? Let us introduce coordinate $\sigma \in [0, \pi]$ which covers the gravity region   - Figure \ref{fig:drawing} (c). It is the counterpart of the coordinate $z$ in the non-gravitational region.
Naively, the integral 
\beq
\label{eq:pholo}
\int_{0}^{\pi} A_\si d\si,
\eeq
is not gauge invariant.
In this Section we explain that (\ref{eq:pholo}) is gauge invariant and so one has to integrate over it. Obviously,
its presence can again be traded for a twisted boundary conditions for the matter fields.

In order to see that, we need to examine the boundary conditions for the gauge field at $\si=0,\pi$ \cite{Harlow:2019yfa}. They should be chosen such that the gauge symmetry actually leads to a global symmetry at the boundary.  Two examples of possible boundary conditions are
\beq
\text{Dirichlet: } A_x \at_\pr = 0, \qquad \text{Neumann: } F_{x \si} \at_\pr = 0.
\eeq
For the Dirichlet boundary conditions it is clear that upon a gauge transformation,
\beq
A_\mu \ra A_\mu + \pr_\mu \alpha,
\eeq
parameter $\alpha$ should not
depend on $x$ at $\si=0,\pi$,
in order to preserve the boundary condition. This way we get two independent physical global symmetries at each of the two boundaries. In other words, non-zero $\alpha(\si=0,\pi)$ corresponds to an actual physical phase rotation which can be measured by an observer. For Neumann boundary conditions we do not have such restrictions. We conclude that in our setup we have to impose Dirichlet boundary conditions.
Since the strip $\si \in [0,\pi]$ is coupled to a flat spacetime region with one global symmetry, we have to restrict to $\alpha(\si=0)=\alpha(\si=\pi)\ \text{mod} \ 2\pi$.
Finally, we can prove that (\ref{eq:pholo}) is gauge-invariant modulo $2\pi$.
Actual gauge transformations have vanishing (again modulo $2\pi$) $\alpha$ parameter at $\si=0, \pi$. Hence, we can only affect (\ref{eq:pholo}) with
\beq
\alpha(\sigma) =  n (\sigma + \pi),\ n \in \mathbb{Z}.
\eeq
Hence the "partial holonomy" (\ref{eq:pholo}) is gauge-invariant modulo $2\pi$.

\section{Avoiding paradox with purity}
\label{sec:paradox}
One can ask if the gravitationally prepared state of Chen, Gorbenko and Maldacena is pure. This question was addressed in the original paper and the answer is positive: $\Tr(\rho^2)=1$. In this Section we would like to argue that the existence of a global symmetry contradicts the purity.
For gauge symmetries the situation is resolved if we integrate over the holonomies in the connected saddle.

When evaluating simple observables, the connected saddle effectively looks thermal. In general, this does not lead to a paradox. However, \textit{assuming} there is a conserved charge this can lead to a paradox. Our discussion here is motivated by a recent paper on fluctuation entropy \cite{Milekhin:2021lmq}.

Suppose the system has a conserved charge $Q$. Here we are talking about the total charge of the system. In our setup $Q$ can be an integral of a conserved current along the whole Cauchy slice, $Q=\oint dx j_0$. \textit{Assume} that the gravitationally prepared bra-ket density matrix $\rho$ commutes with $Q$:
\beq
[Q,\rho]=0.
\eeq
If we do not insert any sources in the path integral, we should expect this equality to hold: the integral over $j_0$ can be freely moved around the cylinder. 
It means that we can diagonalize both $Q$ and $\rho$ at the same time and $\rho$ can be split into blocks of definite charge:
\beq
\rho = \oplus p(q) \rho_q,\ \sum_q p(q) = 1, \Tr(\rho_q)=1.
\eeq
The von Neumann entropy has the following decomposition 
\cite{Belin:2013uta,PhysRevLett.120.200602, Capizzi:2020jed, Murciano:2020vgh, Bonsignori:2020laa}
:
\beq
\label{eq:qresolv}
S(\rho) = \sum_q p(q) S(\rho_q) - \sum_q p(q) \log p(q).
\eeq
Since $S(\rho_q) \ge 0$, the state has a chance to be pure($S(\rho)$ vanishes) only if
$p(q)$ is non-zero for one specific charge.
However, it is easy to see that
$p(q)$ can be found from evaluating a simple observable
\beq
\bra  e^{i Q \alpha} \ket = \sum_q p(q) e^{i \alpha q}.
\eeq
If it has a thermal expectation value then obviously $p(q)$ is non-zero for many charges and the state cannot be pure.

This is yet another reincarnation of the statement "no global symmetries in quantum gravity".  How is the paradox resolved for a gauge symmetry? The decomposition (\ref{eq:qresolv}) still holds in this case. The resolution is that the integration over the holonomies prohibit the charge to flow through the wormhole.
It means that $e^{i Q \alpha}$ will only measure zero charge.

\section{Casimir energy and holonomies}
\label{sec:casimir}
The $U(N)$ symmetry arises if we have $N$ complex Dirac (both chiralities) fermions $\psi_i, i=1,\dots,N$.
In this case once fermions go around the circle they can be twisted by $U(N)$ matrix $U$. We have to integrate over $U$.
We can decompose the twist matrix $U$
as $U=V^\dagger D V$, with $D$ having only diagonal entries $e^{i \theta_i}, \theta_i \in [-\pi,\pi]$. Integration over $V$ will produce a non-trivial measure (Haar measure) for $\theta_i$. 
Now, Haar measure for $U(N)$ is simply the Vandermonde determinant \cite{Aharony:2003sx}:
\beq
\mu_{\rm Haar} = \sum_{i<j} \l \sin \l \frac{\th_i -\th_j}{2} \r \r^2.
\eeq
Notice that once we move this expression to exponent, it will give us a term of order $N^2$.
This is what we call Casimir entropy.

Now lets discuss Casimir energy. Suppose the circle has length $d$.
We can move the matrix $V$ inside the fermions and we obtain the following (diagonal) boundary conditions determined by $D$:
\beq
\psi_i(0) = e^{i (\th_i+\pi)} \psi_i(d), \quad \th_i \in [-\pi,\pi]
\eeq
We have shifted all $\th_i$ by $\pi$ for further convenience.
This leads to the following spectrum for $\psi_i$ excitations (absolute value takes care of "electrons and holes"):
\beq
E_{i,n} = \frac{2\pi}{d} | n + \frac{1}{2} + \frac{\th_i}{2\pi} |
\eeq
In order to compute the Casimir energy we need to properly regularize the sum over the positive-energy modes:
\beq
E_{\rm Casimir} = \lim_{\ep \ra 0} \sum_{n,i} E_{i,n} e^{-\ep E_{i,n}}
\eeq
The answer is 
\beq
\label{eq:Casimir}
E_{\rm Casimir} = -\frac{\pi}{6 d} \sum_{i=1}^N \l 1 - 12 \l \frac{\th_i}{2\pi} \r^2 \r
\eeq
As we should have expected, the energy is minimal for fermions having antiperiodic boundary conditions $\th_i=0$. This energy is of order $N$. 
The point is that Casimir entropy is of order $N^2$, so it might dominate over the energy term.

Also it is clear that the energy and the Haar measure push the eigenvalues $\theta_i$ in different directions. The energy is minimal when all $\theta_i=0$. However this is a measure zero configuration. Naively, the measure factor is of order $N^2$, whereas the energy is only of order $N$. Imagine for a moment that we neglect the influence of energy and try to maximize the measure. Haar measure pushes $\theta_i$ apart. It is well--known (e.g. \cite{Aharony:2003sx}) that the minimum action (maximal measure) configuration for $\theta_i$ is a uniform distribution on $[-\pi,\pi]$:
$\th_i = -\pi + 2 \pi i/N$, because they repel each other. 
It means that in the large $N$ limit the resulting Casimir energy is simply given by the integral over $\th$: 
\beq
\label{eq:zeroC}
\int_{-\pi}^{\pi} d\th \l 1 - 12 \l \frac{\th}{2\pi} \r^2 \r = 0,
\eeq
which is zero! So backreaction from the Casimir energy term will be very important.

\section{Bra-ket wormhole action}
\label{sec:braket}
Let us start from describing the full gravitational action for the bra-ket wormhole without extra symmetries. In this section, we first review the original bra-ket wormhole setup \cite{Chen:2020tes}. \ We then compute and discuss the properties of bra-ket wormholes sourced by matter with extra symmetries. 

\subsection{Bra-ket setup}

As shown in Figure \ref{fig1}, there are disconnected and connected solutions contributing to the gravitational path integrals.  Both solutions have Euclidean $AdS_2$ geometry that is glued to a non-gravitational flat regions. The non-gravitational region has a compact spatial direction $x \sim x +L$.

The action in the gravitational region is the Jackiw--Teitelboim 
(JT) gravity plus matter field,
\begin{align}
    I= -\f{S_0}{4\pi} \l \int R + 2\int K \r - \f{1}{4\pi} \l \int \phi(R+2) + 2\phi_b \int K \r + I_{\rm matter}.
\end{align}
Here we assume the matter is a two-dimensional CFT with a central charge $N$.
At the interface, $\left. \phi\right|_{\rm bdy} = \phi_b = \f{\phi_r}{\epsilon}$ and $\epsilon$ is assumed to be small. 

The disconnected solution is topologically a cylinder. The solution for the metric and dilaton for the flat region is given by \cite{Chen:2020tes}
\begin{align}
& ds^2 = \f{dz^2+dx^2}{\epsilon^2},\qquad \phi = \phi_r/\epsilon,\qquad |z|<\tau,
\end{align}
and for the gravity region, we have two solutions corresponding to top and the bottom in Figure \ref{fig1} (b), each of the form
\begin{align}
ds^2 =  \f{dx^2 + d\sigma^2 }{\sinh^2(2\pi\sigma/L)},\qquad \phi = \f{2\pi}{L}\f{\phi_r}{-\tanh(2\pi\sigma/L)} , \qquad \sigma\le -\epsilon.
\end{align}
In general, the matter stress tensor changes the dilaton profile from the pure JT solution. However, for the disconnected solution, the stress-tensor due to Weyl anomaly on the hyperbolic disk exactly cancels the Casimir stress tensor on the cylinder.

Nonetheless, the matter do play a key role in the connected solution. The metric and dilaton in the flat region are the same as above. In the $AdS_2$ region, the metric is 
\begin{align}
    ds^2 = \f{d\chi^2 + d\sigma^2}{\sin^2(\sigma)},\qquad \chi \sim \chi+b, \qquad -\sigma_c \le \sigma \le \pi-\sigma_c,
\end{align}
where $\chi$ and $x$ are related by a scaling, i.e. $b = \f{\sigma_c}{\epsilon} L$. The dilaton solution depends on the backreacted stress tensor. For translationally-invariant states, the matter stress energy can be written in terms of the matter partition function. However, in the limit where the wormhole throat $b$ is large, the torus becomes like a cylinder. Therefore, the stress tensor of a CFT can be approximated by the Casimir energy on a cylinder \cite{Chen:2020tes}. Variable $\sigma_c$ is related to the UV cutoff  and is small \cite{Chen:2020tes}, so the effective length of the other cycle in the $\sigma$ coordinate is 
\begin{align}
d= \pi + \f{2\tau b}{L}
\end{align}

The throat length $b$ in this limit can be found by comparing different contributions to the action. JT gravity has the following action on this configuration:
\beq
S_{\rm grav} = \frac{\phi_r b^2}{2 \pi L},
\eeq
where $\phi_r$ is the renormalized dilaton value of JT gravity.
Assuming that $b$ is large, CFT action will be dominated by vacuum state propagating along $x$ direction:
\beq
S_{CFT} = b E_{\rm Casimir}.
\eeq

On top of that we also have Weyl anomaly factor because we have $AdS_2$ region:
\beq
S_{\rm anomaly} = \frac{b N}{24}.
\eeq
In total we have
\beq
Z= \int db \exp \l -\frac{\phi_r b^2}{2 \pi L} - \frac{b N}{24} - b E_{\rm Casimir}  \r.
\eeq
We see that it is crucial to have negative Casimir energy in order to have a saddle-point in $b$.

\subsection{Wormholes sourced by gauge symmetries}
Now we turn to the specific setup when $U(N)$ symmetry is gauged. Casimir energy has the following form:
\beq
E_{\rm Casimir} = -\frac{1}{6 \l 1 + \frac{2 \tau b}{\pi L} \r} \sum_i \l 1 - 12 \l \frac{\th_i}{2 \pi} \r^2 \r.
\eeq
This is the analogue of eq. (\ref{eq:Casimir}). 
In total, we have
\beq
Z = \int \  db d\th_1 \dots d\th_N e^{-S},
\eeq
\beq
S = \frac{\phi_r b^2}{2 \pi L} + \frac{bN}{24} - f(b) \sum^N_{i=1} \l 1 - 12 \l \frac{\th_i}{2 \pi} \r^2 \r - \sum^N_{i \neq j} \log \sin \l \frac{|\th_i - \th_j|}{2} \r,
\eeq
with
\beq
f(b)=\frac{b}{6 \l 1 + \frac{2 \tau b}{\pi L} \r}.
\eeq

We can put aside the integral over $b$ and just discuss the integral over $\theta$. Casimir energy term provides a global quadratic potential pushing $\theta_i$ towards $0$. Haar measure provides repulsion. As we saw in the previous Section, for uniformly spread $\theta_i$ Casimir energy is zero.
It is very bad, as it has to be negative enough to overcome the conformal anomaly term, which effectively looks like positive energy. It means that the coefficient $f(b)$, which effectively controls the magnitude of Casimir energy, has to be big enough in order for bra-ket wormhole to exist. In order to compete with the Haar measure, it has to be at least of order $N$. In Appendix \ref{app:int} we find that we need
\beq
\frac{f(b)}{N} > 0.27.
\eeq
Because of the specific form of $f(b)$, it has a maximum as a function of $b$ if $\tau$ is fixed. It leads to the following bound:
\beq
\label{eq:time}
\frac{N \tau}{L} < 0.1.
\eeq
The full phase diagram obtained semi-analytically in Appendix \ref{app:int} is shown in Figure \ref{fig:phase}. It also shows a minimal value for $L/\phi_r$:
\beq
\label{eq:space}
\frac{L}{\phi_r} \gtrsim 33.
\eeq

One interesting feature of the quadratic attractive potential is that even for very small $f(b)$, $\theta_i$ are located within an interval $[-\th_c,\th_c]$ inside $[-\pi,\pi]$. In the language of Yang--Mills matrix models we are always above the Hagedorn phase transition.

\begin{figure}
    \centering
    \minipage{0.47\textwidth}
    \includegraphics[scale=0.41]{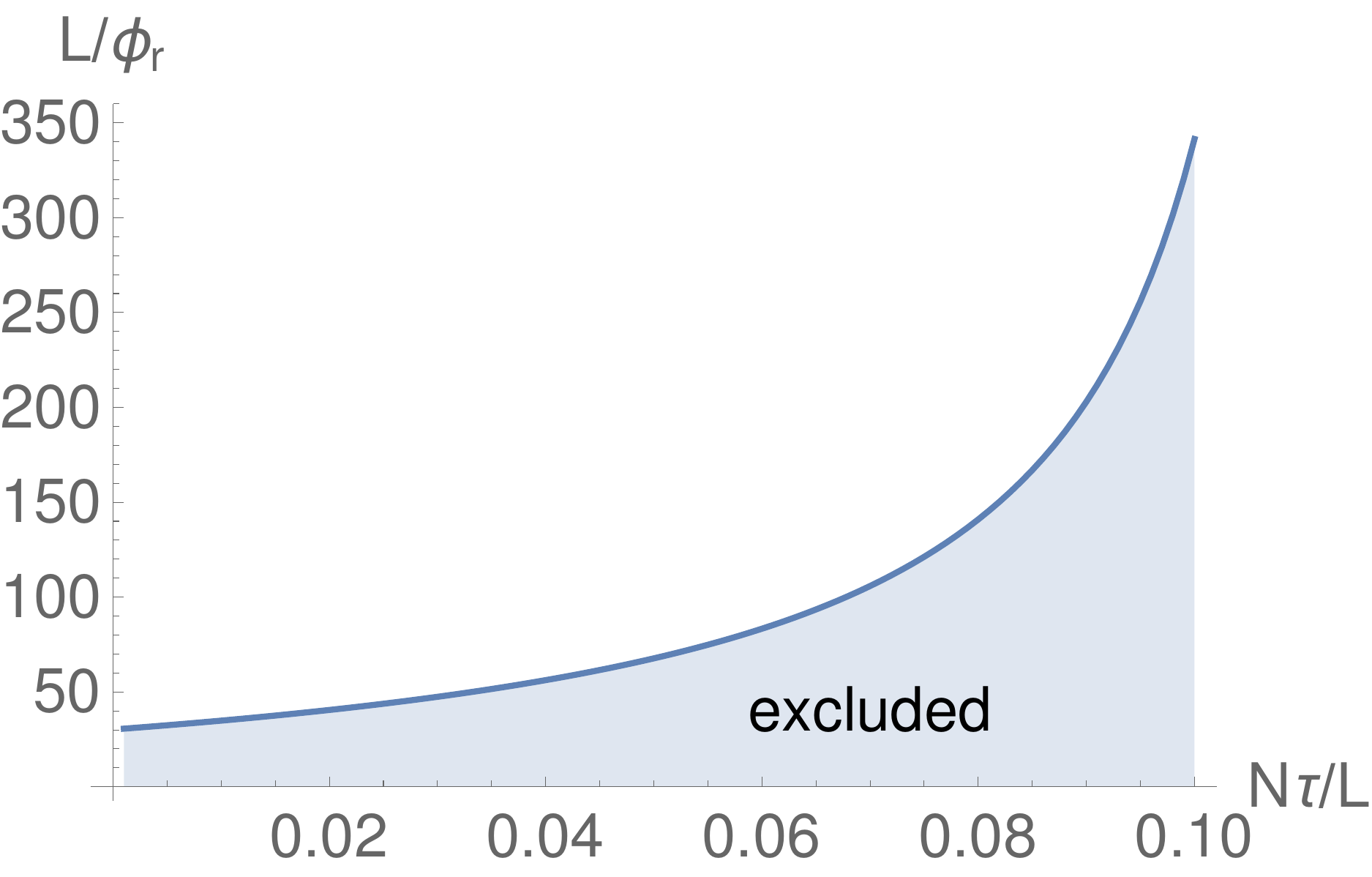}
    \endminipage
    \minipage{0.47\textwidth}
    \includegraphics[scale=0.41]{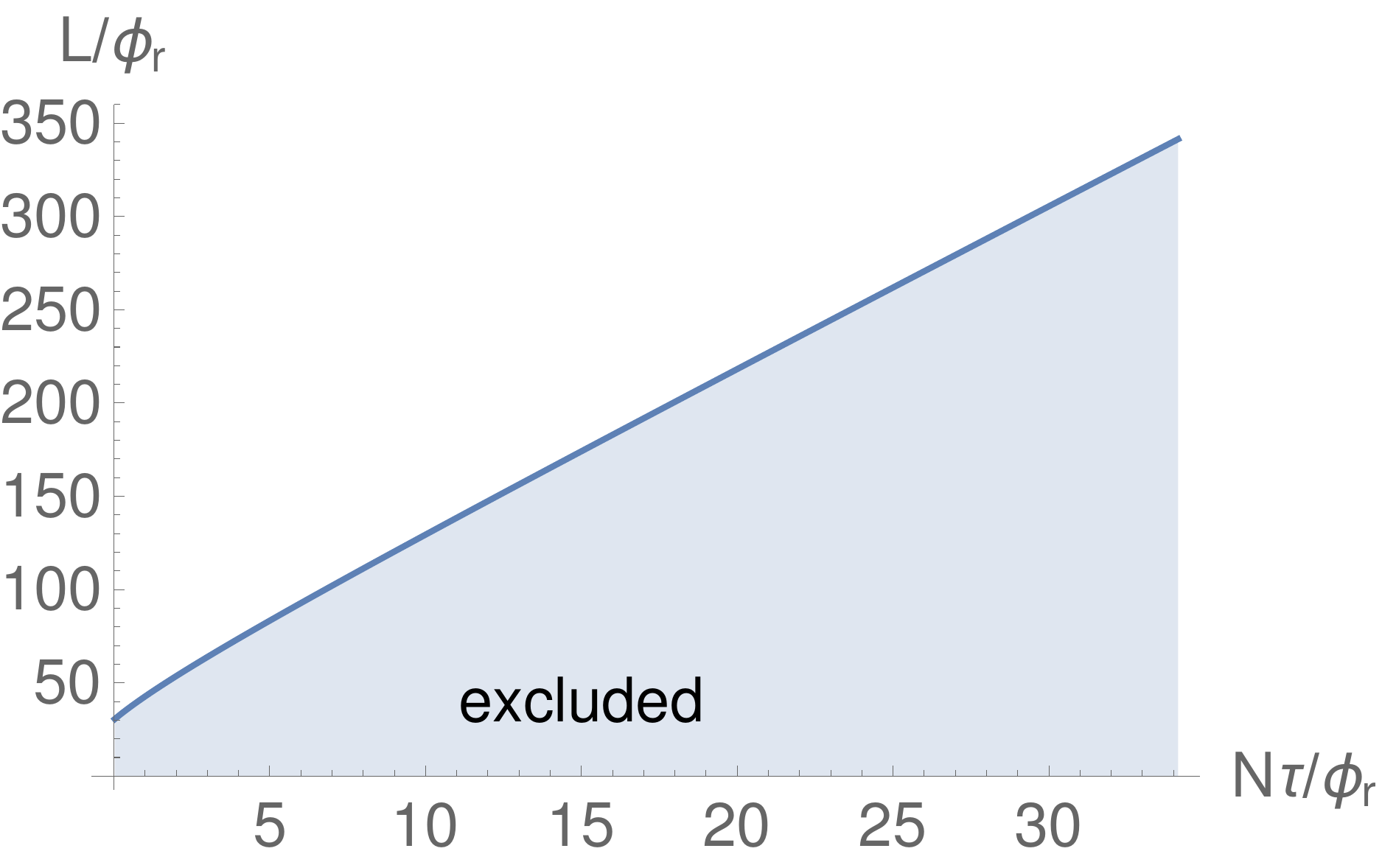}
    \endminipage
    
    \caption{The diagram indicating when bra-ket wormhole can exist. Both plots represent the same phase diagram but drawn using different coordinates. Without Casimir entropy term all parameters are allowed.}
    \label{fig:phase}
\end{figure}

\section{SSA paradox}
\label{sec:ssa}
It turns out that in a certain regime of $L, \phi_r, S_0, N$ two-disk saddle( Figure \ref{fig:drawing} (b))  is pathological: matter entanglement entropy violates SSA, when one computes it using the island prescription \cite{Chen:2020tes}. The resolution is that it is actually the bra-ket wormhole saddle that dominates in this regime. The main result of the previous Section is that it is harder for the bra-ket wormhole to dominate because of Casimir entropy. Hence we would like to check that we do not run into any paradoxes.

We should start the discussion by pointing out that the presence of gauge symmetries should modify the island prescription. This issue was discussed in detail in \cite{Milekhin:2021lmq}. Unfortunately, it is hard to compute entanglement entropy in this case, even for zero gauge coupling and in the OPE limit. However, it was argued in \cite{Milekhin:2021lmq} that the correction is only logarithmic, so it should not affect the violation of SSA.

\begin{figure}
    \centering
    \includegraphics{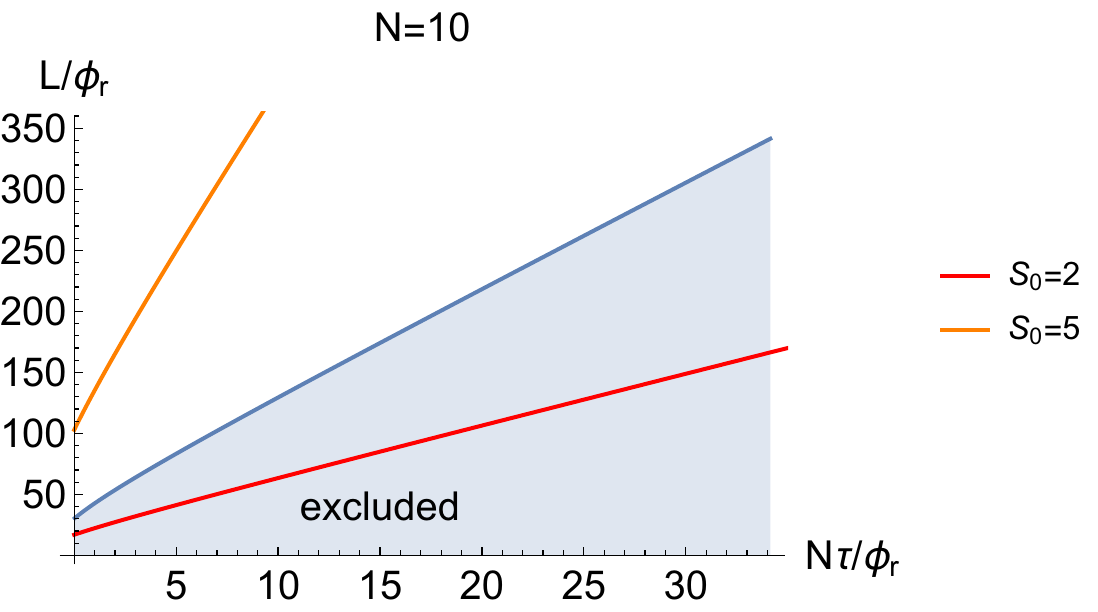}
    \caption{A numerical comparison between when bra-ket wormhole can exist (above blue line) and when it needs to exist (above orange and red lines).  }
    \label{fig:comp}
\end{figure}

It was estimated in \cite{Chen:2020tes} that SSA is violated when an island start to dominate in the two-disk geometry. For $\tau \ll \phi_r/N$ it happens when $L$ is large enough:
\beq
\label{eq:Lmint0}
\frac{L}{\phi_r} \ge \frac{6 \pi}{N} \exp \l \frac{6 S_0}{N}+1 \r.
\eeq
In another regime, $\phi_r/N \ll \tau \ll L$ one needs
\beq
L \ge 4 \pi \tau \exp \l \frac{6 S_0}{N} + \frac{6 \phi_r}{N \tau}  \r.
\eeq
Comparing this to the bounds (\ref{eq:time}),(\ref{eq:space}) we see that we need to require
\beq
\label{eq:s0}
S_0 \gtrsim \frac{1}{6} N \log N,
\eeq
otherwise there is an SSA paradox, but bra-ket wormhole cannot exist. 
For intermediate $\tau$ one has to check numerically that the paradox does not arise as long as the inequality (\ref{eq:s0}) is satisfied. A sample plot for $N=10$ is shown in Figure \ref{fig:comp}.

Strictly speaking, we also have to check that the bra-ket wormhole will dominate over the two-disk saddle. Generically, we will be interested in the region above the blue line in Figures \ref{fig:phase} and \ref{fig:comp}. In this regime the factor $\frac{1}{N} \sum_i \l 1-12 \l \frac{\theta_i}{2\pi} \r^2 \r<1$ in the Casimir energy (\ref{eq:Casimir}) is actually of order $1$ so it affects the action by an order $1$ amount which turns out to be unimportant. For example, for $\tau=0$, bra-ket wormhole will dominate when
\beq
L/\phi_r \ge \# \frac{1}{N} \frac{S_0}{N},
\eeq
where $\#$ is the aforementioned factor of order $1$. Comparing this to the inequality (\ref{eq:Lmint0}) we find that the wormhole will indeed dominate as long as $S_0 \gg N$ regardless of $\#$.
\section*{Acknowledgment}
We would like to thank S.~Giddings, M.~Maldacena, D.~Marolf, H.~Maxfield, M.~Mezei, J.~Wu for discussions and especially E.~Colafranceschi for comments and carefully reading the manuscript. 
The work of A.M. is supported by the Air Force Office of Scientific Research under 
award number FA9550-19-1-0360. It was also supported in part by funds from the University of California. A.M. gratefully acknowledge support from the Simons Center for Geometry and Physics, Stony Brook University at which some of the research for this paper was performed. The work of A.T. is supported in part by a grant from the
Simons foundation and in part by funds from the University of California.
\appendix

\section{Integrating over the eigenvalues}
\label{app:int}
In this Appendix we discuss the evaluation of the following integral:
\beq
Z = \int \  db d\th_1 \dots d\th_N e^{-S},
\eeq
\beq
S = \frac{\phi_r b^2}{2 \pi L} + \frac{bN}{24} - f(b) \sum_i \l 1 - 12 \l \frac{\th_i}{2 \pi} \r^2 \r - \sum_{i \neq j} \log \sin \l \frac{|\th_i - \th_j|}{2} \r,
\eeq
in the large $N$ limit. 
In the bra-ket wormhole setup 
\beq
f(b)=\frac{b}{6 \l 1 + \frac{2 \tau b}{\pi L} \r}.
\eeq

Let us present the main physical picture:
in order to have a bra-ket wormhole we must find a saddle-point with $b>0$.
It is only possible if the Casimir energy $-\sum_i \l 1 - 12 \l \frac{\th_i}{2 \pi} \r^2 \r$ is negative enough to overcome \textit{positive} energy contribution $b N/24$ from the Weyl anomaly.
Haar term $\log \sin(|\th_i-\th_j|)$ want to push $\th_i$ apart, making the Casimir energy smaller in absolute value, whereas $1 - 12 \l \frac{\th_i}{2 \pi} \r^2$ attract them to $\th_i=0$ making the Casimir energy more negative. Obviously, the later effect should win. We will see that it puts a lower bound on $f(b)$.  

Introducing the eigenvalue density $\rho(\th)$,
\beq
\rho(\th) = \frac{1}{N} \sum_i \rho(\th-\th_i),
\eeq
we can write down the action as 
\beq
S = \frac{\phi_r b^2}{2 \pi L} + \frac{bN}{24} - N f(b) + \frac{3N}{\pi^2} f(b) \int d\th \th^2 \rho(\th)  - N^2 \int d\th d\th' \rho(\th) \rho(\th') \log \sin \l \frac{|\th - \th'|}{2} \r. 
\eeq
The integral over $\rho$ looks quadratic.
We can decompose it into Fourier modes:
\beq
\rho = \frac{1}{2 \pi} + \frac{1}{\pi} \sum_{n=1}^{+\infty} \rho_n \cos(n \th)
\eeq
The normalization condition is satisfied automatically, whereas the positivity constraint
is very non-trivial in the $\rho_n$ basis.
If $\rho$ is supported on the whole $[-\pi,\pi]$, then we can treat $\rho_n$ as independent variables. We see that the extremum occurs 
at
\beq
\rho_n = -\frac{6 f(b) (-1)^n}{n N}
\eeq
Which corresponds to the following density $\rho$:
\beq
\rho(\theta) = \frac{1}{2 \pi} + \frac{3 f(b)}{N} \log \l 2 + 2 \cos(\th) \r
\eeq
This is a bad distribution because it is negative for $\th$ close to $\pm \pi$. It means that the actual saddle point $\rho$ is non-zero
only within a finite interval $[-\th_c,\th_c] \subset [-\pi,\pi]$.

Saddle-point equations look like:
\beq
\label{eq:saddle1}
\frac{\phi_r b}{\pi L} = -\frac{N}{24} + N f'(b) \bra 1- \frac{3}{\pi^2} \th^2 \ket,
\eeq
\beq
\l 1-\frac{3}{\pi^2} \th^2 \r \frac{f(b)}{N} = \int d\th' \rho(\th') \log \sin^2 \frac{|\th-\th'|}{2} + \la.
\eeq
Extra $\la$ comes from inserting a Lagrange multiplier for $\int d \th \rho(\theta)=1$ and the expectation value $\bra \ket$ is with respect to $\rho(\theta)$.

\begin{figure}
    \centering
    \minipage{0.47\textwidth}
    \includegraphics[scale=0.35]{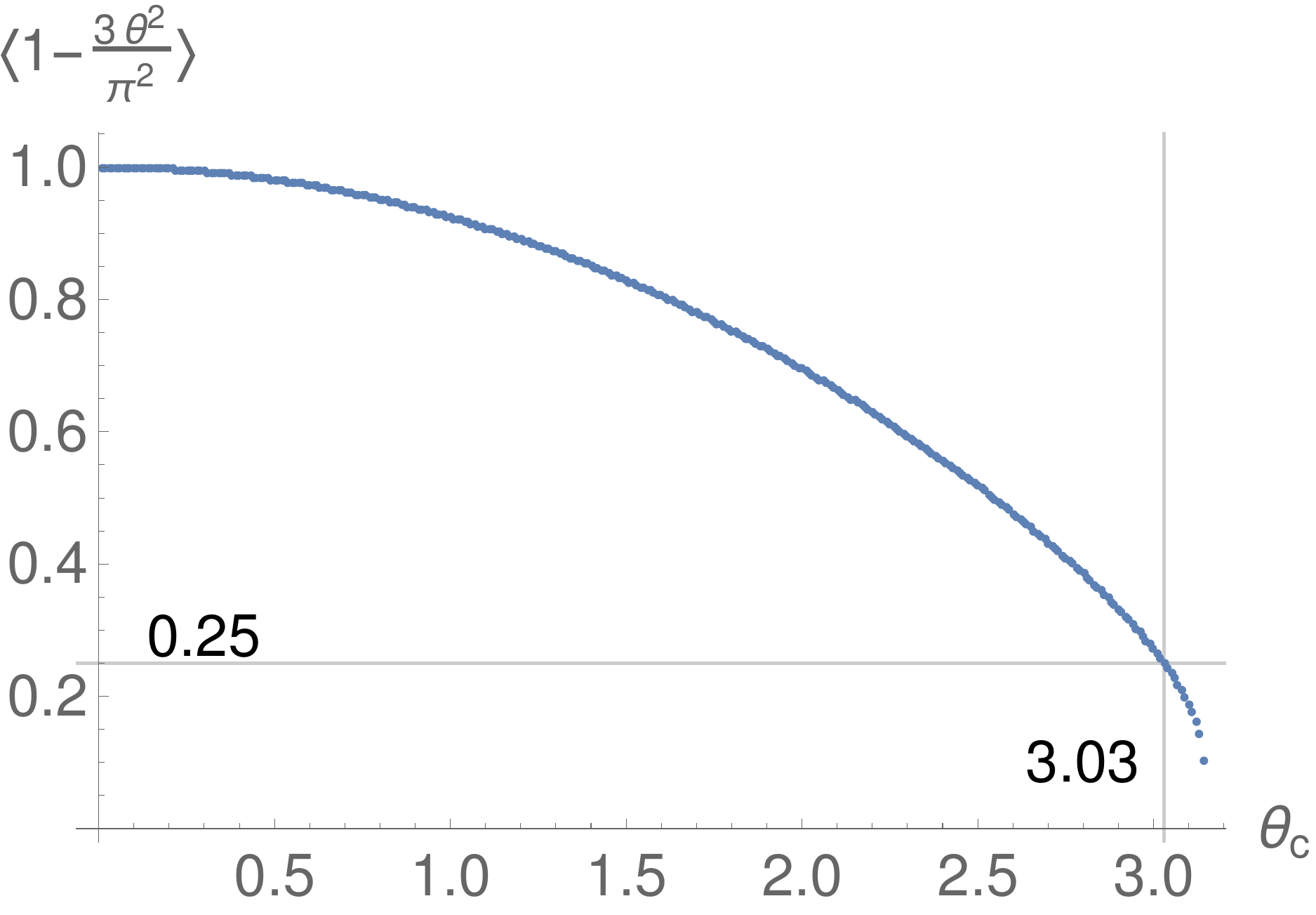}
    \endminipage
    \minipage{0.47\textwidth}
    \includegraphics[scale=0.35]{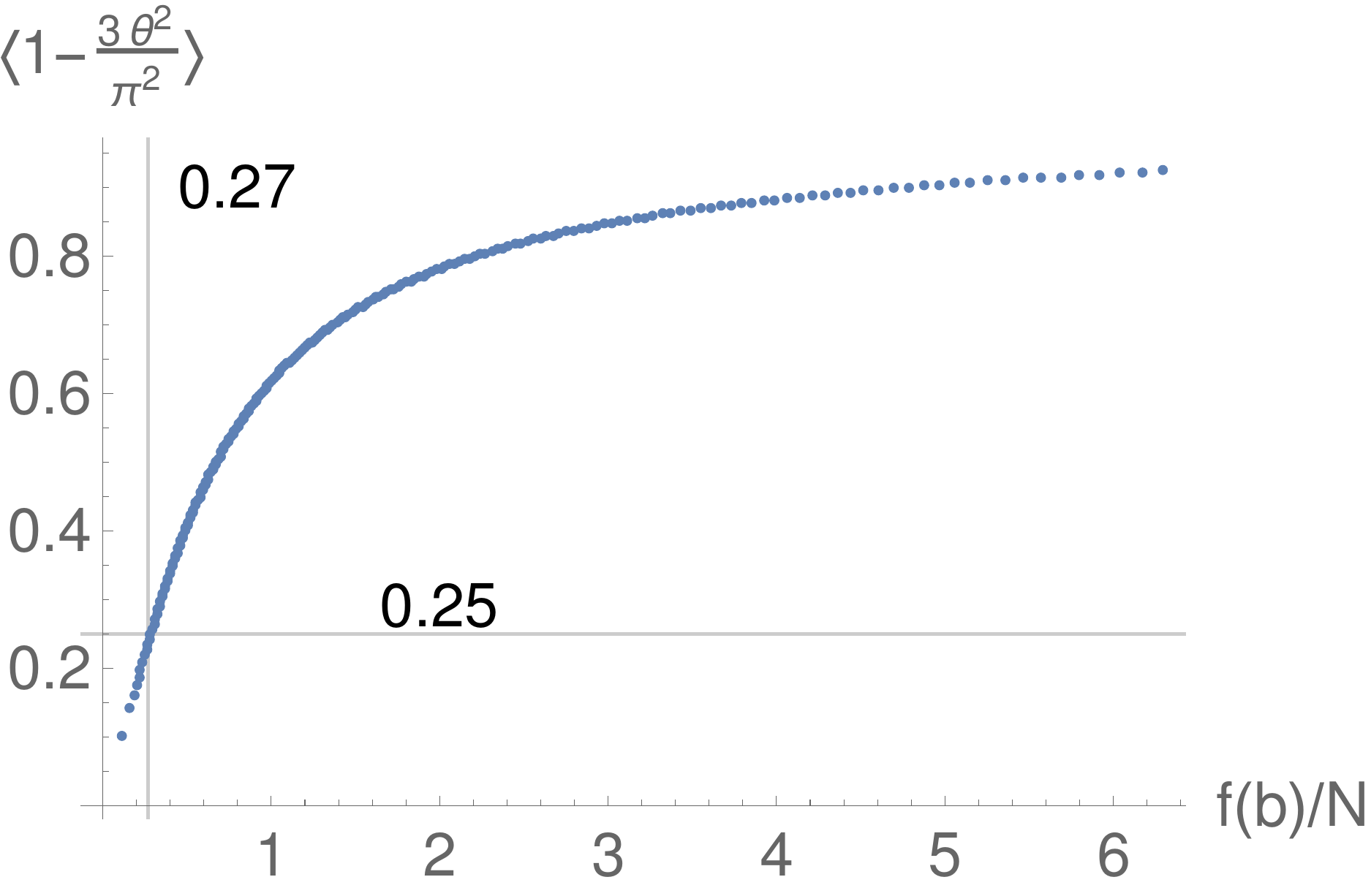}
    \endminipage
    
    \caption{Values of $\bra 1-3 \th^2/\pi^2 \ket$ and $f(b)/N$ obtained numerically using the procedure described in this Section with $n_{max}=200$. Very close to $\th_c = \pi$ the convergence is poor, but luckily we are not interested in the details there. }
    \label{fig:plots}
\end{figure}

Let us concentrate on the second equation first. We know that the solution has a finite support on some unknown interval $[-\th_c,\th_c]$. In fact the whole solution can be parametrized by $\th_c$.
This problem was solved in full generality in  \cite{Aharony:2003sx,Zalewski}. Suppose we are interested in a more general problem with a potential determined by the Fourier coefficients $a_n$: 
\beq
\sum_{n=1} \frac{2 a_n}{n} \cos(n \th)= \int d\th' \rho(\th') \log \sin^2 \frac{|\th-\th'|}{2} + \la.
\eeq
Then the solution is given by :
\beq
\label{eq:rho_sol}
\rho(\th) = \frac{1}{\pi}
\sqrt{\sin^2 \frac{\th_c}{2}-\sin^2 \frac{\th}{2}} \sum_{n=1}^{+\infty} Q_n \cos \l (n-1/2) \th \r,
\eeq
with
\beq
Q_n = 2 \sum^{+\infty}_{l=0} a_{n+l} P_l(\cos(\th_c)),
\eeq
and finally $\th_c$ is determined by
\beq
\label{eq:QQ}
Q_1 = Q_0 +2 .
\eeq
In our case, Fourier coefficients $a_n$ are given by
\beq
a_n = -\frac{f(b)}{N} \frac{6(-1)^n}{n \pi^2}.
\eeq
We see that the solution is determined by a single variable $\th_c$. Instead of setting the potential parameters and solving a transcendental equation (\ref{eq:QQ}), it is better to interpret it in the opposite direction: suppose $\th_c$ is known, then eq. (\ref{eq:QQ}) fixes the overall coefficient in the potential, as $Q_n$ are linear in the potential. Even with $\th_c$ it is still difficult to compute the sum over $n$ in eq. (\ref{eq:rho_sol}) and evaluate the expectation value $\bra 1-3 \th^2/\pi^2 \ket$. 
Therefore we adopt the following semi-analytical strategy:
\begin{itemize}
    \item Fix $\th_c$.
    \item Define rescaled $\tilde{a}_n$ as $a_n = \f{f(b)}{N}\tilde{a}_n$
    and evaluate the corresponding $\tilde{Q}_n$. In fact, one can give \footnote{This can be obtained by taking the generating function for Legendre polynomials, $\sum_{l=0}^{+\infty} t^l P_l(x) = 1/\sqrt{1-2 t x + t^2}$, multiplying by an appropriate power of $t$ and integrating over $t$.} an explicit expression using Appel F1 function(\verb|AppelF1| in Mathematica):
    \beq
    \tilde{Q}_0 = -\frac{12}{\pi^2} \l \log(2) - \log\l 1+ \cos(\theta_c) + \sqrt{2 + 2 \cos(\theta_c)} \r \r 
    \eeq
    \beq
    \tilde{Q}_n = -\frac{12(-1)^n}{\pi^2 n}
    \text{AppelF1} \l n, \f{1}{2},\f{1}{2}, n+1, -e^{i \theta_c}, -e^{-i \theta_c}\r,\ n \ge 1.
    \eeq
    They have the following asymptotics at large $n$, which can be obtained from the Gauss integral representation for the Appel function:
    \beq
    \tilde{Q}_n \approx -\frac{24}{\pi^2 \cos(\theta_c)} \frac{(-1)^n}{n},\ n \gg 1 .
    \eeq
    \item Truncate the sum over $n$ in the full solution (\ref{eq:rho_sol}) to some $n=n_{max}$ to find $\rho$. Here it is convenient to subtract the asymptotics of $\widetilde{Q}_n$ to have better convergence.
    Numerically find $\bra 1-3 \th^2/\pi^2 \ket$.
    
    \item Use evaluated $\Qtt_{0,1}$ and  $\bra 1-3 \th^2/\pi^2 \ket$ to solve the saddle point equations.
    
    To get rid of the explicit factors of $N$, we can define $\tilde{b}=b/N$ and $\tilde{\tau}=\tau N$.
This way we get:
\beq
\frac{1}{6} \frac{\tilde{b}}{1+\frac{2 \tilde{b} \tilde{\tau}}{\pi L}} = \frac{2}{\tilde{Q_1}-\tilde{Q_0}}, \quad \text{(from definition of $f(b)$)}
\label{eq:1}
\eeq
\beq
\frac{\phi_r \tilde{b}}{\pi L} = -\frac{1}{24} + \frac{\bra 1-3 \th^2/\pi^2 \ket}{6 \l 1+ \frac{2 \tilde{t} \tilde{b}}{\pi L} \r^2}.
\label{eq:2}
\eeq
    
    \item Repeat this procedure for all $\th_c \in [-\pi,\pi]$. This way we get Figure \ref{fig:plots}.
\end{itemize}

Even without doing any numerics, we could have noticed that for all actual physical solutions $\bra 1-3 \th^2/\pi^2 \ket$ should be greater than $1/4$, to have a positive value of $b$. It means that the Casimir potential coefficient $f(b)$ has to be big enough. From numerics:
\beq
\boxed{\frac{f(b)}{N} > 0.27} .
\eeq
However with a finite Euclidean time propagation $\tau$, function $f(b)/N$ has a maximum(as a function of $b$): 
\beq
\max_b \frac{1}{6} \frac{\tilde{b}}{1+\frac{2 \tilde{b} \tilde{\tau}}{\pi L}} = \frac{\pi L}{12 \tilde{\tau}}.
\eeq
It implies that at least $N \tau/L \le 0.96$. The actual numerical result is
\beq
\boxed{ \frac{N \tau}{L} < 0.1}.
\eeq
These results were previously states in the main text and illustrated by Figure \ref{fig:phase}.
Finally, the fact that there is a minimum value of $L$ even at $\tau=0$ can also be understood semi-analytically:
$b$ is uniquely determined by $\th_c$ from the first saddle-point eq. (\ref{eq:1}). 
For $\th_c$ close to 0, $b$ has to be very large. But from the second eq. (\ref{eq:2}), the ratio $\phi_r b/L$ has to stay of order 1. Hence $L$ has to be large. In the other regime, when $\th_c$ approaches $3.03$, where $\bra 1-3\th^2/\pi^2 \ket$ approaches $1/4$, $b$ stays finite. But at the same time the ratio $\phi_r b/L$ has to be very small, because the right hand side of eq. (\ref{eq:2}) vanishes. So $L$ has to grow again. Somewhere in the middle it reaches a minimal value.
%%%%%%%%%%%%%%%%%%%%%%%%%%%%%%%%%%%%%%%%%%%%%%%%%%%%%%%%
%\begin{thebibliography}{1}
%\end{thebibliography}
%\printbibliography

\bibliographystyle{ourbst.bst}
\bibliography{references.bib}
\end{document}